\documentclass{article}
\usepackage{amssymb}
\usepackage{amsfonts}
\usepackage{amsmath}

\newtheorem{theorem}{Theorem}
\newtheorem{acknowledgement}[theorem]{Acknowledgement}

\newtheorem{corollary}[theorem]{Corollary}

\newtheorem{lemma}[theorem]{Lemma}

\newtheorem{proposition}[theorem]{Proposition}
\newtheorem{remark}[theorem]{Remark}

\newenvironment{proof}[1][Proof]{\noindent\textbf{#1.} }{\ \rule{0.5em}{0.5em}}
\input{tcilatex}

\begin{document}

\title{Squeezed Coherent States and a Semiclassical Propagator for the Schr%
\"{o}dinger Equation in Phase Space}
\author{Maurice A de Gosson \\
Universit\"{a}t Potsdam, Inst. f. Mathematik \\
Am Neuen Palais 10, D-14415 Potsdam\\
E-mail address: maurice.degosson@gmail.com \and Serge M de Gosson \\
V\"{a}xj\"{o} Universitet, MSI\\
SE-351 95 V\"{a}xj\"{o}, Sweden\\
E-mail address: sergedegosson@gmail.com}
\maketitle

\begin{abstract}
We construct semiclassical solutions of the symplectically covariant Schr%
\"{o}dinger phase-space equation rigorously studied in a previous paper; we
use for this purpose an adaptation of Littlejohn's nearby-orbit method. We
take the opportunity to discuss in some detail the so fruitful notion of
squeezed coherent state and the action of the metaplectic group on these
states.
\end{abstract}

\section{Introduction}

Let $\psi $ be a square-integrable solution of Schr\"{o}dinger's equation 
\begin{equation*}
i\hslash \frac{\partial \psi }{\partial t}=H(x,-i\hbar \partial _{x})\psi
\end{equation*}%
and set $\Psi =U_{\phi }\psi $, where the \emph{wave-packet transform }$%
U_{\phi }$ is defined as follows \cite{physa}: 
\begin{equation}
U_{\phi }\psi (x,p)=\left( \tfrac{1}{2\pi \hbar }\right) ^{n/2}e^{\frac{i}{%
2\hbar }p\cdot x}\int e^{-\frac{i}{\hbar }p\cdot x^{\prime }}\psi (x^{\prime
})\overline{\phi }(x-x^{\prime })d^{n}x^{\prime }  \label{ufi}
\end{equation}%
here $\phi $ is an \emph{arbitrary} rapidly decreasing function. The
function $\Psi $ satisfies the equation 
\begin{equation}
i\hslash \frac{\partial \Psi }{\partial t}=\widehat{H}_{\text{ph}}\Psi
\label{sph}
\end{equation}%
where the operator $\widehat{H}_{\text{ph}}$ is defined by the intertwining
formula 
\begin{equation*}
U_{\phi }\widehat{H}_{\text{ph}}=H(x,-i\hbar \partial _{x})U_{\phi };
\end{equation*}%
A straightforward calculation shows that we have%
\begin{equation}
U_{\phi }(x\psi )=(\tfrac{1}{2}x+i\hbar \partial _{p})U_{\phi }\psi \text{ \
, \ }U_{\phi }(-i\hbar \partial _{x}\psi )=(\tfrac{1}{2}p-i\hbar \partial
_{x})U_{\phi }\psi ;  \label{13}
\end{equation}%
these relations motivate the notation 
\begin{equation*}
\widehat{H}_{\text{ph}}=H(\tfrac{1}{2}x+i\hbar \partial _{p},\tfrac{1}{2}%
p_{j}-i\hbar \partial _{x})
\end{equation*}%
and we may thus rewrite (\ref{sph}) as the \emph{phase-space Schr\"{o}dinger
equation}%
\begin{equation}
i\hbar \frac{\partial }{\partial t}\Psi =H(\tfrac{1}{2}x+i\hbar \partial
_{p},\tfrac{1}{2}p_{j}-i\hbar \partial _{x})\Psi .  \label{syco}
\end{equation}%
In a recent paper \cite{physa} (also see \cite{phystv}) one of us exposed in
some detail the properties of the solutions of this equation, originally
proposed by Torres-Vega and Frederick \cite{TV1,TV2} in the particular case
where $\phi $ is a Gaussian.

In this paper we pursue the study of the properties of the solutions of this
equation, and, in particular we show how to construct in an easy way
approximate solutions to that equation for a large class of initial wave
functions. Our approach is an adaptation to phase space of Littlejohn's \cite%
{Littlejohn} \emph{nearby orbit method }(we remark that Heller \cite{Heller}
has considered a more restrictive notion).

\subsection*{Notation}

The position vector will be denoted by $x=(x_{1},...,x_{n})$ and the
momentum vector by $p=(p_{1},...,p_{n})$, and we write $z=(x,p)$ for the
generic phase space variable. We will use the generalized gradients%
\begin{equation*}
\partial _{x}=\left( \tfrac{\partial }{\partial x_{1}},...,\tfrac{\partial }{%
\partial x_{n}}\right) \text{ , }\partial _{p}=\left( \tfrac{\partial }{%
\partial p_{1}},...,\tfrac{\partial }{\partial p_{n}}\right)
\end{equation*}%
and $\partial _{z}=(\partial _{x},\partial _{p})$.

The symplectic product of $z=(x,p)$, $z^{\prime }=(x^{\prime },p^{\prime })$
is denoted by $\sigma (z,z^{\prime })$; by definition:%
\begin{equation*}
\sigma (z,z^{\prime })=p\cdot x^{\prime }-p^{\prime }\cdot x
\end{equation*}%
where the dot $\cdot $ is the usual (Euclidean) scalar product. In matrix
notation:%
\begin{equation*}
\sigma (z,z^{\prime })=(z^{\prime })^{T}Jz\text{ \ , \ }J=%
\begin{pmatrix}
0_{n\times n} & I_{n\times n} \\ 
-I_{n\times n} & 0_{n\times n}%
\end{pmatrix}%
\text{.}
\end{equation*}%
The corresponding symplectic group is denoted by $\func{Sp}(n)$: the
relation $S\in \func{Sp}(n)$ means that $S$ is a real $2n\times 2n$ matrix
such that $\sigma (Sz,Sz^{\prime })=\sigma (z,z^{\prime })$; equivalently $%
S^{T}JS=SJS^{T}=J$.

For $z_{0}=(x_{0},p_{0})$ the Heisenberg--Weyl operator $\widehat{T}(z_{0})$
acts on functions of $x$ via 
\begin{equation}
\widehat{T}(z_{0})\psi (x)=e^{\frac{i}{\hbar }(p_{0}\cdot x-\frac{1}{2}%
p_{0}\cdot x_{0})}\psi (x-x_{0}).  \label{HW}
\end{equation}%
The Wigner--Moyal transform of a pair $(\psi ,\phi )$ of functions in the
Schwartz space $\mathcal{S}(\mathbb{R}_{x}^{n})$ is defined by%
\begin{equation}
W(\psi ,\phi )(z)=\left( \tfrac{1}{2\pi \hbar }\right) ^{n}\int e^{-\tfrac{i%
}{\hbar }p\cdot y}\psi (x+\tfrac{1}{2}y)\overline{\phi (x-\tfrac{1}{2}y)}%
d^{n}y\text{.}  \label{wimo}
\end{equation}

\section{Coherent States in Phase Space\label{cohst}}

In this Section we review known results about coherent states; the action of
the metaplectic group on these states will be studied in the next section.

We denote by $\phi ^{\hbar }$ the standard coherent state:%
\begin{equation}
\phi ^{\hbar }(x)=\left( \tfrac{1}{\pi \hbar }\right) ^{n/4}e^{-\frac{1}{%
2\hbar }x^{2}};  \label{a}
\end{equation}%
for $z_{0}=(x_{0},p_{0})$ we define more generally the coherent state $\phi
_{z_{0}}^{\hbar }=\widehat{T}(z_{0})\phi _{\hbar }$ that is%
\begin{equation}
\phi _{z_{0}}^{\hbar }(x)=e^{\frac{i}{\hbar }(p_{0}\cdot x-\frac{1}{2}%
p_{0}\cdot x_{0})}\phi ^{\hbar }(x-x_{0}).  \label{b}
\end{equation}%
(These states are often denoted respectively by $\left\vert 0\right\rangle $
and $\left\vert z_{0}\right\rangle $ in bra-ket notation \cite{Littlejohn}).
The following properties are well-known (see for instance. \cite{KS,Pere};
for the sake of self-containedness we however give a proof.

\begin{proposition}
\label{propcoh}(i) We have%
\begin{equation}
\left( \tfrac{1}{2\pi \hbar }\right) ^{n}\int \phi _{z_{0}}^{\hbar }(x)%
\overline{\phi _{z_{0}}^{\hbar }(y)}d^{2n}z_{0}=\delta (x-y).  \label{ortho}
\end{equation}%
(ii) For every $\psi \in L^{2}(\mathbb{R}^{n})$ 
\begin{equation}
\psi (x)=\left( \tfrac{1}{2\pi \hbar }\right) ^{n}\int (\psi ,\phi
_{z_{0}}^{\hbar })_{L^{2}}\phi _{z_{0}}^{\hbar }(x)d^{2n}z_{0}
\label{Fourier}
\end{equation}%
and%
\begin{equation}
||\psi ||_{L^{2}}^{2}=\left( \tfrac{1}{2\pi \hbar }\right) ^{n}\int |(\psi
,\phi _{z_{0}}^{\hbar })_{L^{2}}|^{2}d^{2n}z_{0}.  \label{foucoff}
\end{equation}%
(iii) Let $\widehat{A}$ be a continuous linear operator $\mathcal{S}(\mathbb{%
R}^{n})\longrightarrow \mathcal{S}^{\prime }(\mathbb{R}^{n})$ with Schwartz
kernel $K$; we have%
\begin{equation}
K(x,y)=\left( \tfrac{1}{2\pi \hbar }\right) ^{n}\int \widehat{A}\phi
_{z_{0}}^{\hbar }(x)\overline{\phi _{z_{0}}^{\hbar }(y)}d^{2n}z_{0}.
\label{kernel}
\end{equation}
\end{proposition}

\subsection{Squeezed coherent states}

Despite the interest of the standard coherent states it is often
advantageous to work with their generalization, called \textquotedblleft
squeezed coherent states\textquotedblright\ (they anyway appear
automatically when one lets the metaplectic group act on coherent states,
see Subsection \ref{subsecmpco}). A squeezed coherent state is a Gaussian\
of the type%
\begin{equation}
\phi _{(X,Y)}^{\hbar }(x)=\left( \tfrac{1}{\pi \hbar }\right) ^{n/4}(\det
X)^{1/4}e^{-\frac{1}{2\hbar }(X+iY)x^{2}}  \label{squ1}
\end{equation}%
where $X$ and $Y$ are real symmetric $n\times n$ matrices with $X>0$; we
have $||\phi _{(X,Y)}^{\hbar }||_{L^{2}}=1$. We will find it convenient to\
set 
\begin{equation*}
M=i(X+iY)\text{ , }X=X^{T}>0\text{ , }Y=Y^{T}.
\end{equation*}%
and to write $\phi _{(X,Y)}^{\hbar }=\phi _{M}^{\hbar }$; thus:%
\begin{equation*}
\phi _{M}^{\hbar }(x)=\left( \tfrac{1}{\pi \hbar }\right) ^{n/4}(\det
X)^{1/4}e^{\frac{i}{2\hbar }Mx^{2}}\text{ \ , \ }X=\func{Im}M.
\end{equation*}

The Wigner transform of $\phi _{M}^{\hbar }$ is given by the formula%
\begin{equation}
W\phi _{M}^{\hbar }(z)=\left( \tfrac{1}{\pi \hbar }\right) ^{n}(\det
X)^{-1/2}e^{-\frac{1}{\hbar }Gz^{2}}  \label{wsqu}
\end{equation}%
where $G$ is the symmetric matrix%
\begin{equation}
G=%
\begin{pmatrix}
X+YX^{-1}Y & YX^{-1} \\ 
X^{-1}Y & X^{-1}%
\end{pmatrix}%
\text{;}  \label{ge}
\end{equation}%
an essential remark is that $G$ is in addition \textit{symplectic }(this
fact was apparently first observed by Bastiaans \cite{Bast}). More
precisely, we have $G=S^{T}S$ with%
\begin{equation}
S=%
\begin{pmatrix}
X^{1/2} & 0 \\ 
X^{-1/2}Y & X^{-1/2}%
\end{pmatrix}%
\in \limfunc{Sp}(n)  \label{ess}
\end{equation}%
as results from a direct calculation; notice that $S$ belongs to the
isotropy subgroup of the Lagrangian plane $\ell _{p}=0\times \mathbb{R}^{n}$
in $\limfunc{Sp}(n)$. For $z_{0}\in \mathbb{R}^{2n}$ we set 
\begin{equation*}
\phi _{z_{0},M}^{\hbar }=\widehat{T}(z_{0})\phi _{M}^{\hbar };
\end{equation*}%
the Wigner transform of $\phi _{z_{0},M}^{\hbar }$ is given by 
\begin{equation}
W\phi _{z_{0},M}^{\hbar }(z)=W\phi _{M}^{\hbar }(z-z_{0});  \label{squ2}
\end{equation}%
it is thus a Gaussian centered at $z_{0}$.

Let us generalize formula (\ref{wsqu}) by calculating the Wigner-Moyal
transform $W(\phi _{z_{0},M}^{\hbar },\phi _{z_{0}^{\prime },M^{\prime
}}^{\hbar })$ of a pair of squeezed coherent states; recall for this purpose
the \textquotedblleft Fresnel formula\textquotedblright 
\begin{equation}
\left( \tfrac{1}{2\pi \hbar }\right) ^{n/2}\int e^{-\frac{i}{\hbar }\xi
\cdot x}e^{-\frac{1}{2\hbar }Kx^{\prime 2}}d^{n}x^{\prime }=(\det
K)^{-1/2}e^{-\frac{1}{2\hbar }K^{-1}\xi ^{2}}  \label{fresnel}
\end{equation}%
valid for $\xi \in \mathbb{C}^{n}$, $K=K^{T}$, $\func{Im}K>0$; here $(\det
K)^{-1/2}=\lambda _{1}^{-1/2}\cdots \lambda _{n}^{-1/2}$ where $\lambda
_{j}^{-1/2}$ is the square root with positive real part of the eigenvalue $%
\lambda _{j}^{-1}$ of $K^{-1}$ (see \cite{Folland,Leray}).

\begin{proposition}
\label{ww}We have%
\begin{equation}
W(\phi _{M}^{\hbar },\phi _{M^{\prime }}^{\hbar })(z)=\left( \tfrac{1}{\pi
\hbar }\right) ^{n}(\det XX^{\prime })^{-1/4}e^{-\frac{1}{\hbar }Fz^{2}}
\label{psps}
\end{equation}%
where $F$ is the matrix%
\begin{equation}
F=%
\begin{pmatrix}
2i\overline{M^{\prime }}(M-\overline{M^{\prime }})^{-1}M & -i(M+\overline{%
M^{\prime }})(M-\overline{M^{\prime }})^{-1} \\ 
-i(M-\overline{M^{\prime }})^{-1}(M+\overline{M^{\prime }}) & 2i(M-\overline{%
M^{\prime }})^{-1}%
\end{pmatrix}%
.  \label{fm}
\end{equation}
\end{proposition}

\begin{proof}
Setting $W_{M,M^{\prime }}=W(\phi _{M}^{\hbar },\phi _{M^{\prime }}^{\hbar
}) $ we have%
\begin{equation*}
W_{M,M^{\prime }}(z)=\left( \tfrac{1}{2\pi \hbar }\right) ^{n}\int e^{-\frac{%
i}{\hbar }py}\phi _{M}^{\hbar }(x+\tfrac{1}{2}y)\overline{\phi _{M^{\prime
}}^{\hbar }}(x-\tfrac{1}{2}y)d^{n}y
\end{equation*}%
that is, replacing $\phi _{M}^{\hbar }$ and $\phi _{M^{\prime }}^{\hbar }$
by their expressions,%
\begin{equation*}
W_{M,M^{\prime }}(z)=C(X,X^{\prime })e^{\frac{i}{2\hbar }(M+\overline{%
M^{\prime }})x^{2}}\int e^{-\frac{i}{\hbar }py}e^{\frac{i}{2\hbar }\Phi
(x,y)}d^{n}y
\end{equation*}%
where $M=i(X+iY)$, $M^{\prime }=i(X^{\prime }+iY^{\prime })$, the terms $%
C(X,Y)$ and $\Phi (x,y)$ being given by%
\begin{align*}
C(X,Y)& =2^{-n}\left( \tfrac{1}{\pi \hbar }\right) ^{2n}(\det XX^{\prime
})^{1/4} \\
\Phi (x,y)& =M(x+\tfrac{1}{2}y)^{2}+\overline{M^{\prime }}(x-\tfrac{1}{2}%
y)^{2}\text{.}
\end{align*}%
Expanding $\Phi (x,y)$ and using (\ref{fresnel}) we finally get%
\begin{equation*}
W_{M,M^{\prime }}(z)=\left( \tfrac{1}{\pi \hbar }\right) ^{n}(\det
XX^{\prime })^{-1/4}e^{-\frac{1}{\hbar }Fz^{2}}
\end{equation*}%
where $F$ is the matrix (\ref{fm}) (the left-upper block is obtained by
using the identity%
\begin{equation*}
M+\overline{M^{\prime }}-(M-\overline{M^{\prime }})(M+\overline{M^{\prime }}%
)^{-1}(M-\overline{M^{\prime }})=4\overline{M^{\prime }}(M+\overline{%
M^{\prime }})^{-1}M\text{. )}
\end{equation*}
\end{proof}

Noting the following formula, which is an easy consequence of the
definitions of the Wigner--Moyal transform and the Heisenberg--Weyl
operators:

\begin{lemma}
For all $z_{0},z_{0}^{\prime }$ in $\mathbb{R}_{z}^{2n}$ and $f,g$ in $L^{2}(%
\mathbb{R}_{z}^{2n})$ we have%
\begin{multline*}
W(\widehat{T}(z_{0})\psi ,\widehat{T}(z_{0}^{\prime })\psi ^{\prime })(z)=e^{%
\frac{i}{\hbar }(\sigma (z_{0}-z_{0}^{\prime },z)-\frac{1}{2}\sigma
(z_{0},z_{0}^{\prime }))} \\
\times W(\psi ,\psi ^{\prime })(z-\tfrac{1}{2}(z_{0}+z_{0}^{\prime }))
\end{multline*}
\end{lemma}

\noindent we immediately get the following generalization of Proposition \ref%
{ww}:

\begin{corollary}
We have%
\begin{equation*}
W(\phi _{z_{0},M}^{\hbar },\phi _{z_{0},M^{\prime }}^{\hbar })(z)=\left( 
\tfrac{1}{\pi \hbar }\right) ^{n}e^{\frac{i}{\hbar }(\sigma
(z_{0}-z_{0}^{\prime },z)-\frac{1}{2}\sigma (z_{0},z_{0}^{\prime }))}(\det
XX^{\prime })^{-1/4}e^{-\frac{1}{\hbar }F(z-\tfrac{1}{2}(z_{0}+z_{0}^{\prime
}))^{2}}
\end{equation*}%
where $F$ is given by (\ref{fm}); hence in particular%
\begin{equation*}
W(\phi _{z_{0},M}^{\hbar },\phi _{z_{0},M^{\prime }}^{\hbar })(z)=\left( 
\tfrac{1}{\pi \hbar }\right) ^{n}(\det XX^{\prime })^{-1/4}e^{-\frac{1}{%
\hbar }F(z-z_{0})^{2}}.
\end{equation*}
\end{corollary}

\subsection{Phase-space coherent states\label{pscs}}

For each $\phi \in \mathcal{S}(\mathbb{R}_{x}^{n})$ the operator 
\begin{equation*}
U_{\phi }\psi (z)=\left( \tfrac{\pi \hbar }{2}\right) ^{n/2}W(\psi ,\phi )(%
\tfrac{1}{2}z)
\end{equation*}%
is an isometry of $L^{2}(\mathbb{R}_{x}^{n})$ onto its range $\mathcal{H}%
_{\phi }$; it follows that:

\begin{proposition}
Let $\Phi _{z_{0}}^{\hbar }=U_{\phi }(\phi _{z_{0}}^{\hbar })$. For each $%
\Psi \in \mathcal{H}_{\phi }$ we have%
\begin{equation}
\Psi (z)=\left( \tfrac{1}{2\pi \hbar }\right) ^{n}\int (\Psi ,\Phi
_{z_{0}}^{\hbar })_{L^{2}}\Phi _{z_{0}}^{\hbar }(z)d^{2n}z_{0}  \label{gdpsi}
\end{equation}%
and%
\begin{equation}
||\Psi ||_{L^{2}}=\left( \tfrac{1}{2\pi \hbar }\right) ^{n}\int |(\Psi ,\Phi
_{z_{0}}^{\hbar })_{L^{2}}|^{2}d^{2n}z_{0}.  \label{gdpsi2}
\end{equation}
\end{proposition}

\begin{proof}
Let $\psi $ be defined by $\Psi =U_{\phi }\psi $; In view of part (ii) of
Proposition \ref{propcoh} (formula (\ref{Fourier})) we have%
\begin{equation*}
\psi =\left( \tfrac{1}{2\pi \hbar }\right) ^{n}\int (\psi ,\phi
_{z_{0}}^{\hbar })_{L^{2}}\phi _{z_{0}}^{\hbar }d^{2n}z_{0}
\end{equation*}%
hence, since $U_{\phi }$ is continuous, 
\begin{eqnarray*}
\Psi &=&\left( \tfrac{1}{2\pi \hbar }\right) ^{n}\int (\psi ,\phi
_{z_{0}}^{\hbar })_{L^{2}}U_{\phi }(\phi _{z_{0}}^{\hbar })d^{2n}z_{0} \\
&=&\left( \tfrac{1}{2\pi \hbar }\right) ^{n}\int (\psi ,\phi _{z_{0}}^{\hbar
})_{L^{2}}\Phi _{z_{0}}^{\hbar }d^{2n}z_{0}.
\end{eqnarray*}%
In view of formula (\ref{parseval}) we have 
\begin{equation*}
(\Psi ,\Phi _{z_{0}}^{\hbar })_{L^{2}}=(U_{\phi }\psi ,U_{\phi }\phi
_{z_{0}}^{\hbar })_{L^{2}}=(\psi ,\phi _{z_{0}}^{\hbar })_{L^{2}}
\end{equation*}%
hence (\ref{gdpsi}); formula (\ref{gdpsi2}) follows by a similar argument
from formula (\ref{foucoff}). \ 
\end{proof}

\section{Metaplectic Group and Coherent States}

The metaplectic group $\limfunc{Mp}(n)$ is a faithful unitary representation
of $\limfunc{Sp}_{2}(n)$, the double cover of the symplectic group $\limfunc{%
Sp}(n)$. There are several different ways to describe the elements of $%
\limfunc{Mp}(n)$ (see for instance \cite{Birk,Leray,Wallach}); for our
purposes the most adequate definition makes use the notion of generating
function for a free symplectic matrix because it is the simplest way to
arrive at the Weyl symbol of metaplectic operators (and hence to their
extension to phase space).

The interest of the metaplectic representation comes from the fact that it
links in a crucial way classical (Hamiltonian) mechanics to quantum
mechanics (see for instance \cite{MF} or \cite{Birk} and the references
therein). Assume in fact that $H$ is a Hamiltonian function which is a
quadratic polynomial in the position and momentum variables (with possibly
time-dependent coefficients): thus%
\begin{equation*}
H(z,t)=\frac{1}{2}z^{T}H^{\prime \prime }(t)z=\frac{1}{2}H^{\prime \prime
}(t)z^{2}
\end{equation*}%
where $H^{\prime \prime }(t)$ is a symmetric matrix (it is the Hessian
matrix of $H$). The associated Hamilton equations $\dot{z}=\partial
_{z}H(z,t)$ determine a (generally time-dependent) flow consisting of
symplectic matrices $S_{t}$. We thus have a continuous path $t\longmapsto
S_{t}$ in the symplectic group $\limfunc{Sp}(n)$ passing through the
identity at time $t=0$: $S_{0}=I$. Following general principles, this path
can be lifted (in a unique way) to a path $t\longmapsto \widehat{S}_{t}$ in $%
\limfunc{Mp}(n)$ such that $\widehat{S}_{0}=\widehat{I}$ (the identity in $%
\limfunc{Mp}(n)$). Choose now an initial wavefunction $\psi _{0}$ in, say, $%
\mathcal{S}(\mathbb{R}_{x}^{n})$ and set $\psi (x,t)=\widehat{S}_{t}\psi
_{0}(x)$. The function $\psi $ satisfies Schr\"{o}dinger's equation%
\begin{equation*}
i\hbar \frac{\partial \psi }{\partial t}=H(x,i\hbar \partial _{x},t)\psi
\end{equation*}%
where the operator $H(x,i\hbar \partial _{x})$ is given by%
\begin{equation*}
H(x,i\hbar \partial _{x},t)=\frac{1}{2}(x,i\hbar \partial _{x})^{T}H^{\prime
\prime }(t)(x,i\hbar \partial _{x}).
\end{equation*}

\subsection{Description of $\limfunc{Mp}(n)$ and $\limfunc{IMp}(n)$}

Let $W$ be quadratic form of the type 
\begin{equation*}
W(x,x^{\prime })=\tfrac{1}{2}Px^{2}-Lx\cdot x^{\prime }+\tfrac{1}{2}Qx^{2}
\end{equation*}%
with $P=P^{T}$, $Q=Q^{"}$, $\det L\neq 0$; (we have set $Px^{2}=Px\cdot x$,
etc.). To each such quadratic form we associate the generalized Fourier
transform%
\begin{equation}
\widehat{S}_{W,m}\psi (x)=\left( \tfrac{1}{2\pi i\hbar }\right) ^{n/2}i^{m}%
\sqrt{|\det L|}\int e^{\frac{i}{\hbar }W(x,x^{\prime })}\psi (x^{\prime
})d^{n}x^{\prime }  \label{swm}
\end{equation}%
where $m$ corresponds to a choice of the argument of $\det L$ modulo $2\pi $%
. One proves (see \cite{AIF}) that\ every $\widehat{S}\in \limfunc{Mp}(n)$
can be written (of course in a non-unique way) as the product of two
operators of the type (\ref{swm}), and that the integer%
\begin{equation}
m(\widehat{S})=m+m^{\prime }-\func{Inert}(P^{\prime }+Q)  \label{ms}
\end{equation}%
is independent modulo $4$ of the factorization $\widehat{S}_{W,m}\widehat{S}%
_{W^{\prime },m^{\prime }}$ of $\widehat{S}$; the class modulo $4$ of $m(%
\widehat{S})$ is the \emph{Maslov index of the metaplectic operator }$%
\widehat{S}$. Since $\limfunc{Mp}(n)$ is a realization of the double cover
of $\limfunc{Sp}(n)$ there exists a natural projection 
\begin{equation*}
\pi ^{\limfunc{Mp}}:\limfunc{Mp}(n)\longrightarrow \limfunc{Sp}(n);
\end{equation*}%
that projection is a 2-to-1 group epimorphism defined by the condition: $%
S_{W}=\pi ^{\limfunc{Mp}}(\widehat{S}_{W,m})$ \emph{is the free symplectic
matrix generated by }$W$, that is:%
\begin{equation*}
(x,p)=S_{W}(x^{\prime },p^{\prime })\Longleftrightarrow \left\{ 
\begin{array}{c}
p=\partial _{x}W(x,x^{\prime }), \\ 
p^{\prime }=-\partial _{x^{\prime }}W(x,x^{\prime }).%
\end{array}%
\right.
\end{equation*}%
Writing $S\in \limfunc{Sp}(n)$ in block form:%
\begin{equation}
S=%
\begin{pmatrix}
A & B \\ 
C & D%
\end{pmatrix}
\label{sabcd}
\end{equation}%
$S$ is free if and only if $\det B\neq 0$, and the generating function $W$
is the quadratic form in $2n$ variables 
\begin{equation}
W(x,x^{\prime })=\tfrac{1}{2}DB^{-1}x^{2}-(B^{-1})^{T}x\cdot x^{\prime }+%
\tfrac{1}{2}B^{-1}Ax^{2}  \label{w}
\end{equation}

We will use the following covariance properties of the metaplectic operators:

\begin{itemize}
\item We have 
\begin{equation}
\widehat{S}\widehat{T}(z_{0})=\widehat{T}(Sz_{0})\widehat{S}  \label{meco}
\end{equation}%
for all $\widehat{S}\in \limfunc{Mp}(n)$ and $z_{0}\in \mathbb{R}^{2n};$

\item If $\widehat{S}\in \limfunc{Mp}(n)$ has projection $S=\pi ^{\limfunc{Mp%
}}(\widehat{S})$ then 
\begin{equation*}
W(\widehat{S}\psi ,\widehat{S}\phi )=W(\psi ,\phi )\circ S^{-1}.
\end{equation*}
\end{itemize}

The group generated by the Weyl--Heisenberg operators $\widehat{T}(z_{0})$
and the operators $\widehat{S}\in \limfunc{Mp}(n)$ is a group of unitary
operators in $L^{2}(\mathbb{R}^{n})$; it is denoted by $\limfunc{IMp}(n)$
and called the \textit{inhomogeneous metaplectic group}.

\subsection{The action of $\limfunc{Mp}(n)$ on coherent states\label%
{subsecmpco}}

To describe the action of $\widehat{S}\in \limfunc{Mp}(n)$ on the squeezed
coherent states $\phi _{z_{0},M}^{\hbar }$ we will need the following Lemma.
Let us denote by $\Sigma (n)$ the Siegel half-space, that is%
\begin{equation*}
\Sigma (n)=\{M:M=M^{T},\func{Im}M>0\}
\end{equation*}%
($M$ is a complex $n\times n$ matrix).

\begin{lemma}
Let $S\in \limfunc{Sp}(n)$ be given by (\ref{sabcd}) and $M\in \Sigma (n)$.
Then $\det (A+BM)\neq 0$, $\det (C+DM)\neq 0$ and%
\begin{equation}
\alpha (S)M=(C+DM)(A+BM)^{-1}\in \Sigma (n)  \label{sieg}
\end{equation}%
(in particular $\alpha (S)M$ is symmetric), and 
\begin{equation}
\alpha (SS^{\prime })M=\alpha (S)\alpha (S^{\prime })M.  \label{ass}
\end{equation}%
The action $\limfunc{Sp}(n)\times \Sigma (n)\longrightarrow \Sigma (n)$
defined by (\ref{sieg}) is transitive: if $M,M^{\prime }\in \Sigma (n)$ then
there exists $S\in \limfunc{Sp}(n)$ such that $M^{\prime }=\alpha (S)M$.
\end{lemma}

For a proof see \cite{Folland,Haas}; also see the preprint \cite{CR} by
Combescure and Robert. Notice that (\ref{ass}) implies that $S\longmapsto
\alpha (S)$ is a true representation of the symplectic group in the Siegel
half-space.

Let $\widehat{S}\in \limfunc{Mp}(n)$ have projection%
\begin{equation*}
S=\pi ^{\limfunc{Mp}}(\widehat{S})=%
\begin{pmatrix}
A & B \\ 
C & D%
\end{pmatrix}%
\end{equation*}%
on $\limfunc{Sp}(n)$; then%
\begin{equation*}
\widehat{S}\phi ^{\hbar }(x)=\left( \frac{1}{\pi \hbar }\right) ^{n/4}\frac{%
i^{m(\widehat{S})}}{\sqrt{\det (A+iB)}}\exp \left[ -\frac{1}{2\hbar }\alpha
(S)x^{2}\right]
\end{equation*}%
where the branch cut of the square root of $\det (A+iB)$ is taken to lie
just under the positive real axis; $m(\widehat{S})$ is the Maslov index (\ref%
{ms}) of $\widehat{S}$. Using this formula $\widehat{S}\phi _{z_{0}}^{\hbar
}(x)$ is easily calculated: since by definition $\phi _{z_{0}}^{\hbar }=%
\widehat{T}(z_{0})\phi ^{\hbar }$ the metaplectic covariance formula (\ref%
{meco}) immediately yields%
\begin{equation*}
\widehat{S}\phi _{z_{0}}^{\hbar }(x)=\widehat{T}(Sz_{0})\widehat{S}\phi
^{\hbar }(x)
\end{equation*}%
(see Littlejohn \cite{Littlejohn} for an explicit formula).

The results above can be generalized to arbitrary squeezed coherent states:

\begin{proposition}
\label{alpha}Let $\phi _{z_{0},M}^{\hbar }$, $M\in \Sigma (n)$, be a
squeezed coherent state and $\widehat{S}\in \limfunc{Mp}(n)$, $S=\pi ^{%
\limfunc{Mp}}(\widehat{S})$. We have%
\begin{equation}
\widehat{S}\phi _{M}^{\hbar }=\phi _{\alpha (S)M}^{\hbar }\text{\ \ , \ }%
\widehat{S}\phi _{z_{0},M}^{\hbar }=\widehat{T}(Sz_{0})\phi _{\alpha
(S)M}^{\hbar }\text{.}  \label{alpham}
\end{equation}
\end{proposition}

\noindent (see for instance \cite{CR,Folland,Littlejohn}).

The Gaussian character of a wavepacket is thus preserved by metaplectic
operators; as a consequence the solution of a Schr\"{o}dinger equation with
Gaussian initial value remains Gaussian when the Hamiltonian operator is
associated to a quadratic Hamiltonian function.

\subsection{The Weyl symbol of a metaplectic operator\label{subweyl}}

For $S\in \limfunc{Sp}(n)$ such that $\det (S-I)\neq 0$ we set 
\begin{equation*}
M_{S}=\frac{1}{2}J(S+I)(S-I)^{-1}
\end{equation*}%
(it is the symplectic Cayley transform of $S$). In \cite{MdGMW} one of us
proved the following result:

\begin{proposition}
(i) If $\widehat{S}_{W,m}$ is such that $S_{W}=\pi ^{\limfunc{Mp}}(\widehat{S%
}_{W,m})$ has no eigenvalue equal to one, then%
\begin{equation}
\widehat{S}_{W,m}\psi (x)=\left( \frac{1}{2\pi \hbar }\right) ^{n}\frac{i^{m-%
\limfunc{Inert}W_{xx}}}{\sqrt{|\det (S-I)|}}\int e^{\frac{i}{2\hbar }%
M_{S}z_{0}^{2}}\widehat{T}(z_{0})\psi (x)\mathrm{d}^{2n}z_{0}  \label{MW1}
\end{equation}%
for $\psi \in S(R_{x}^{n})$; here $\limfunc{Inert}W_{xx}$ is the number of
negative eigenvalues of the Hessian matrix of the function $x\longmapsto
W(x,x)$. (ii) Every $\widehat{S}\in \limfunc{Mp}(n)$ can be written as the
product of two metaplectic operators of this type. (iii) More generally,
every $\widehat{S}\in \limfunc{Mp}(n)$ with projection $S$ such that $\det
(S-I)\neq 0$ can be written as%
\begin{equation}
\widehat{S}\psi (x)=\left( \frac{1}{2\pi \hbar }\right) ^{n}\frac{i^{\nu (%
\widehat{S})}}{\sqrt{|\det (S-I)|}}\int e^{\frac{i}{2\hbar }M_{S}z_{0}^{2}}%
\widehat{T}(z_{0})\psi (x)\mathrm{d}^{2n}z_{0}.  \label{MW2}
\end{equation}%
where $\nu (\widehat{S})$ is the Conley--Zehnder index of $\widehat{S}$.
\end{proposition}

Formulae (\ref{MW1}) and (\ref{MW2}) yield the Weyl representations of the
metaplectic operators; for a definition and a precise study of the
Conley--Zehnder index $\nu (\widehat{S})$ appearing in (\ref{MW2}) see \cite%
{Birk}. In \cite{MdGMW} it was also proven that formula (\ref{MW2}) can be
rewritten alternatively as 
\begin{equation}
\widehat{S}=\left( \frac{1}{2\pi \hbar }\right) ^{n}i^{\nu (\widehat{S})}%
\sqrt{|\det (S-I)|}\int e^{-\frac{i}{2}\sigma (Sz,z)}\widehat{T}((S-I)z)%
\mathrm{d}^{2n}z  \label{MWa}
\end{equation}%
or 
\begin{equation}
\widehat{S}=\left( \frac{1}{2\pi \hbar }\right) ^{n}i^{\nu (\widehat{S})}%
\sqrt{|\det (S-I)|}\int \widehat{T}(Sz)\widehat{T}(-z)\mathrm{d}^{2n}z\text{.%
}  \label{MWb}
\end{equation}

\section{The Nearby Orbit Method\label{secnom}}

We now have the background material that is needed to study the nearby-orbit
method for the phase-space Schr\"{o}dinger equation. We begin by describing
this method in the usual case; we refer to Littlejohn's excellent review 
\cite{Littlejohn} for a discussion and interpretation.

\subsection{The \textquotedblleft standard\textquotedblright\ case}

Consider an arbitrary (possibly time-dependent) Hamiltonian function $H$ on $%
\mathbb{R}_{z}^{2n}$; we denote by $(f_{t})$ its flow: $t\longmapsto
f_{t}(z_{0})$ is the solution of Hamilton's equations $\dot{z}=J\partial
_{z}H(z,t)$ passing through the phase-space point $z_{0}$ at time $t=0$. We
have, for every $z$,%
\begin{equation*}
f_{t}(z)=f_{t}(z_{0})+S_{t}(z_{0})(z-z_{0})+O((z-z_{0})^{2})
\end{equation*}%
where $S_{t}(z_{0})=Df_{t}(z_{0})$ is the Jacobian matrix of $f_{t}$ at $%
z_{0}$. Since Hamiltonian flows consists of canonical transformations, $%
S_{t}(z_{0})$ is a symplectic matrix: $S_{t}(z_{0})\in \limfunc{Sp}(n)$.
Suppose now the point $z$ is close to $z_{0}$; then%
\begin{equation*}
f_{t}(z)\approx f_{t}(z_{0})+S_{t}(z_{0})(z-z_{0});
\end{equation*}%
the nearby orbit approximation consists in replacing $f_{t}(z)$ by $%
f_{t}(z_{0})+S_{t}(z_{0})(z-z_{0})$. Noticing that 
\begin{equation*}
f_{t}(z_{0})+S_{t}(z_{0})(z-z_{0})=T(f_{t}(z_{0}))S_{t}(z_{0})T(z_{0})^{-1}z
\end{equation*}%
where $T(z_{0})$ is the translation operator $z\longmapsto z+z_{0}$, an
educated guess is that we will obtain the quantum counterpart of this
approximation by replacing translations by Weyl--Heisenberg operators, and
symplectic matrices by metaplectic operators. In fact this guess is correct
up to a phase factor arising from the non-commutativity of the
Heisenberg--Weyl operators: the \textquotedblleft true\textquotedblright\
semiclassical approximation to the solution $\psi $ of Schr\"{o}dinger's
equation%
\begin{equation}
i\hslash \frac{\partial \psi }{\partial t}=\widehat{H}\psi \text{ \ ,\ \ }%
\psi (t=0)=\psi _{0}  \label{S1}
\end{equation}%
is given by 
\begin{equation}
U(t,z_{0})\psi _{0}=e^{\frac{i}{\hbar }\gamma (z_{0},t)}\widehat{T}%
(f_{t}(z_{0}))\widehat{S}_{t}(z_{0})\widehat{T}(z_{0})^{-1}\psi _{0}
\label{nbo1}
\end{equation}%
where the phase $\gamma (t)$ is given by 
\begin{equation}
\gamma (z_{0},t)=\int_{0}^{t}\tfrac{1}{2}\sigma (z(t^{\prime }),\dot{z}%
(t^{\prime }))-H(z(t^{\prime }),t^{\prime })dt^{\prime }.  \label{nbo2}
\end{equation}%
One shows (\cite{CR1,Littlejohn}) that the function $\psi =U(t,z_{0})\psi
_{0}$ is the solution of the Schr\"{o}dinger equation associated to the
Hamiltonian function%
\begin{multline*}
H_{z_{0}}(z,t)=H(f_{t}(z_{0}),t) \\
+H^{\prime }(f_{t}(z_{0}),t)(z-f_{t}(z_{0}))+\frac{1}{2}H^{\prime \prime
}(f_{t}(z_{0}),t)(z-f_{t}(z_{0}))^{2}
\end{multline*}%
obtained by truncating the Taylor series for $H$ at $f_{t}(z_{0})$ after
second order. The approximation (\ref{nbo1}) is valid when the initial state 
$\psi _{0}$ is localized near $z_{0}=(x_{0},p_{0})$, that is when when $%
x_{0} $ and $p_{0}$ are chosen to be the position and momentum expectation
vectors $(x\psi _{0},\psi _{0})_{L^{2}}$ and $p_{0}=(p\psi _{0},\psi
_{0})_{L^{2}}$ at initial time $t=0$; the metaplectic operators $\widehat{S}%
_{t}(z_{0})$ are obtained by the usual lifting-procedure: the function $%
t\longmapsto S_{t}(z_{0})$ is a path in the symplectic group passing through
the identity of $\limfunc{Sp}(n)$ at time $t=0$; the function $t\longmapsto 
\widehat{S}_{t}(z_{0})$ is then the path in the metaplectic group passing
through the identity at time $t=0$, and such that $\pi ^{\limfunc{Mp}}(%
\widehat{S}_{t}(z_{0}))=S_{t}(z_{0})$ for all $t$. Formula (\ref{nbo1}) has
the following interpretation (Littlejohn \cite{Littlejohn}): the operator $%
\widehat{T}(z_{0})^{-1}$ rigidly translates the initial wavepacket to a
wavepacket centered at the origin; the metaplectic operator $\widehat{S}%
_{t}(z_{0})$ then \textquotedblleft squeezes\textquotedblright\ it, and $%
\widehat{T}(f_{t}(z_{0}))$ finally rigidly translates this squeezed
wavepacket so that it becomes centered at $f_{t}(z_{0})$; one then
multipliers the result by $e^{\frac{i}{\hbar }\gamma (z_{0},t)}$.

Let us work out explicitly formula (\ref{nbo1}) when the initial wavepacket $%
\psi _{0}$ is a squeezed coherent state:

\begin{proposition}
Let $\phi _{z_{0},M}^{\hbar }$ be an arbitrary squeezed coherent state. We
have:%
\begin{equation*}
U(t,z_{0})\phi _{z_{0},M}^{\hbar }=e^{\frac{i}{\hbar }\gamma (z_{0},t)}\phi
_{f_{t}(z_{0}),\alpha (S_{t}(z_{0}))M}^{\hbar }
\end{equation*}%
where $\alpha (S_{t}(z_{0}))$ is defined by (\ref{sieg}) with $%
S=S_{t}(z_{0}) $.
\end{proposition}

\begin{proof}
By definition $\phi _{z_{0},M}^{\hbar }=\widehat{T}(z_{0})\phi _{M}^{\hbar }$
hence%
\begin{equation*}
U(t,z_{0})\phi _{z_{0},M}^{\hbar }(x)=e^{\frac{i}{\hbar }\gamma (z_{0},t)}%
\widehat{T}(f_{t}(z_{0}))\widehat{S}_{t}(z_{0})\phi _{M}^{\hbar }
\end{equation*}%
that is, taking the first formula (\ref{alpham}) into account,%
\begin{eqnarray*}
U(t,z_{0})\phi _{z_{0},M}^{\hbar } &=&e^{\frac{i}{\hbar }\gamma (z_{0},t)}%
\widehat{T}(f_{t}(z_{0}))\phi _{\alpha (S_{t}(z_{0}))M}^{\hbar } \\
&=&e^{\frac{i}{\hbar }\gamma (z_{0},t)}\phi _{f_{t}(z_{0}),\alpha
(S_{t}(z_{0}))M}^{\hbar }
\end{eqnarray*}%
which completes the proof.
\end{proof}

Formula (\ref{nbo1}) can be extended to arbitrary square-integrable initial
wavefunctions using the machinery of coherent states reviewed in Section \ref%
{cohst}: applying formula (\ref{Fourier})\ to $\psi _{0}\in L^{2}(\mathbb{R}%
_{x}^{n})$ one has%
\begin{equation*}
\psi _{0}(x)=\left( \tfrac{1}{2\pi \hbar }\right) ^{n}\int (\psi _{0},\phi
_{z_{0}}^{\hbar })_{L^{2}}\phi _{z_{0}}^{\hbar }(x)d^{2n}z_{0}
\end{equation*}%
and one then takes as semiclassical solution%
\begin{equation*}
U(t)\psi _{0}(x)=\left( \tfrac{1}{2\pi \hbar }\right) ^{n}\int (\psi
_{0},\phi _{z_{0}}^{\hbar })_{L^{2}}U(t,z_{0})\phi _{z_{0}}^{\hbar
}(x)d^{2n}z_{0}.
\end{equation*}%
Since by definition $\widehat{T}(z_{0})^{-1}\phi _{z_{0}}^{\hbar }=\phi
^{\hbar }$ is the standard coherent state, this formula can be rewritten%
\begin{equation}
U(t)\psi _{0}(x)=\left( \tfrac{1}{2\pi \hbar }\right) ^{n}\int \left( \psi
_{0},\phi _{z_{0}}^{\hbar }\right) _{L^{2}}\widehat{T}(f_{t}(z_{0}))e^{\frac{%
i}{\hbar }\gamma (t)}\widehat{S}_{t}(z_{0}))\phi ^{\hbar }(x)d^{2n}z_{0}
\label{nbo3}
\end{equation}%
(\textit{cf.} equation (7.33) in \cite{Littlejohn}).

\begin{remark}
Observe that neither $U(t,z_{0})$ nor $U(t)$ are linear operators.
\end{remark}

Of course, a natural question is arising at this point, namely
\textquotedblleft How good are the semiclassical approximations (\ref{nbo1}%
), (\ref{nbo2})?\textquotedblright . Very precise estimates have been given
in \cite{CR1,Hag1,Hag2} (see the discussion in the last section); we will
content us here with quoting the following result:

\begin{proposition}
\label{probou}Let $\psi $ be the exact solution to Schr\"{o}dinger's
equation (\ref{S1}) with initial datum $\psi _{0}=\phi _{z_{0}}^{\hbar }$.
Then for each $T>0$ there exists $C_{T}\geq 0$ such that%
\begin{equation}
||U(t,z_{0})\phi _{z_{0}}^{\hbar }-\psi (\cdot ,t)||_{L^{2}(\mathbb{R}%
_{x}^{n})}\leq C_{T}\sqrt{\hbar }  \label{estim}
\end{equation}%
provided that the Hamiltonian function $H$ satisfies uniform estimates of
the type%
\begin{equation}
|\partial _{z}^{\alpha }H(z,t)|\leq C_{\alpha ,T}^{\prime }(1+|z|)^{m}
\label{condh}
\end{equation}%
for $|\alpha |\geq m$, $|t|<T$ and $z\in \mathbb{R}^{2n}$.
\end{proposition}

\noindent (It is in fact possible to obtain precise bounds for the constant $%
C_{T}$; see \cite{CR1}).

\subsection{Weyl operators on phase space}

In \cite{physa} (also see \cite{Birk} for details) we noticed that the
wave-packet transform (\ref{ufi}) is related to the Wigner--Moyal transform (%
\ref{wimo}) by the simple formula 
\begin{equation}
U_{\phi }\psi (z)=\left( \tfrac{\pi \hbar }{2}\right) ^{n/2}W(\psi ,\phi )(%
\tfrac{1}{2}z)\text{ \ , \ }\psi \in L^{2}(\mathbb{R}_{x}^{n})  \label{12}
\end{equation}%
It follows that we have%
\begin{equation}
(U_{\phi }\psi ,U_{\phi }\psi ^{\prime })_{L^{2}(\mathbb{R}_{z}^{2n})}=(\psi
,\psi ^{\prime })_{L^{2}(\mathbb{R}_{x}^{n})}  \label{parseval}
\end{equation}%
and hence each of the linear mappings $U_{\phi }$ is an isometry of $L^{2}(%
\mathbb{R}_{x}^{n})$ onto a \emph{closed} subspace $\mathcal{H}_{\phi }$ of $%
L^{2}(\mathbb{R}_{z}^{2n})$ (the square integrable functions on phase
space). It follows that $U_{\phi }^{\ast }U_{\phi }$ is the identity
operator on $L^{2}(\mathbb{R}_{x}^{n})$ and that $P_{\phi }=U_{\phi }U_{\phi
}^{\ast }$ is the orthogonal projection onto the Hilbert space $\mathcal{H}%
_{\phi }$. Defining $\widehat{T}_{\text{ph}}(z_{0})$ by%
\begin{equation}
\widehat{T}_{\text{ph}}(z_{0})\Psi (z)=e^{-\frac{i}{\hbar }\sigma
(z_{0},z)}\Psi (z-z_{0})  \label{hwp}
\end{equation}%
we have the fundamental relation%
\begin{equation}
\widehat{T}_{\text{ph}}(z_{0})U_{\phi }=U_{\phi }\widehat{T}(z_{0}).
\label{14}
\end{equation}%
Formula (\ref{14}) allows us to associate to every Weyl operator 
\begin{equation*}
\widehat{A}\psi (x)=\left( \tfrac{1}{2\pi \hbar }\right) ^{n}\int a_{\sigma
}(z_{0})\widehat{T}(z_{0})\psi (x)d^{2n}z_{0}
\end{equation*}%
the phase space operator%
\begin{equation}
\widehat{A}_{\text{ph}}\Psi (z)=\left( \tfrac{1}{2\pi \hbar }\right)
^{n}\int a_{\sigma }(z_{0})\widehat{T}_{\text{ph}}(z_{0})\Psi (z)d^{2n}z_{0};
\label{15}
\end{equation}%
of course%
\begin{equation}
\widehat{A}_{\text{ph}}U_{\phi }=U_{\phi }\widehat{A}.  \label{16}
\end{equation}%
Formulae (\ref{15}) and (\ref{MW1}) allow us to give an explicit description
of the action of the metaplectic representation on phase-space functions: if 
$\det (S-I)\neq 0$ we define 
\begin{equation}
\widehat{S}_{\text{ph}}\Psi (z)=\left( \frac{1}{2\pi \hbar }\right) ^{n/2}%
\frac{i^{\nu (S)}}{\sqrt{|\det (S-I)|}}\int e^{\frac{i}{2\hbar }%
M_{S}z_{0}^{2}}\widehat{T}_{\text{ph}}(z_{0})\Psi (z)\mathrm{d}^{2n}z_{0}%
\text{;}  \label{meta}
\end{equation}%
the operators $\widehat{S}_{\text{ph}}$ are in one-to-one correspondence
with the metaplectic operators $\widehat{S}$ and thus generate a group which
we denote by $Mp_{\text{ph}}(n)$; that group is of course isomorphic to $%
\limfunc{Mp}(n)$. Of course, as an immediate consequence of The following
equivalent formulae are immediate consequences of the corresponding
expressions (\ref{MWa}) and (\ref{MWb}) of $\widehat{S}$: 
\begin{equation}
\widehat{S}_{\text{ph}}=\left( \tfrac{1}{2\pi }\right) ^{n}i^{\nu (S)}\sqrt{%
|\det (S-I)|}\int e^{-\frac{i}{2}\sigma (Sz,z)}\widehat{T}_{\text{ph}%
}((S-I)z)\mathrm{d}^{2n}z  \label{alf2}
\end{equation}%
and 
\begin{equation}
\widehat{S}_{\text{ph}}=\left( \tfrac{1}{2\pi }\right) ^{n}i^{\nu (S)}\sqrt{%
|\det (S-I)|}\int \widehat{T}_{\text{ph}}(Sz)\widehat{T}_{\text{ph}}(-z)%
\mathrm{d}^{2n}z\text{.}  \label{alf1}
\end{equation}

Notice that the well-known \textquotedblleft metaplectic
covariance\textquotedblright\ relation $\widehat{A\circ S}=\widehat{S}^{-1}%
\widehat{A}\widehat{S}$ valid for any $\widehat{S}\in \limfunc{Mp}(n)$ with
projection $S\in Sp(n)$ extends to the phase-space Weyl operators $\widehat{A%
}_{\text{ph}}$: we have%
\begin{equation}
\widehat{S}_{\text{ph}}\widehat{T}_{\text{ph}}(z_{0})\widehat{S}_{\text{ph}%
}^{-1}=\widehat{T}_{\text{ph}}(Sz)\text{ \ , \ }\widehat{A\circ S}_{\text{ph}%
}=\widehat{S}_{\text{ph}}^{-1}\widehat{A}_{\text{ph}}\widehat{S}_{\text{ph}}.
\label{metaco}
\end{equation}

In Subsection \ref{pscs} we defined coherent states in phase space. The
metaplectic action on coherent states described in Proposition \ref{alpha}
carries over to this case without difficulty, yielding the formulae%
\begin{equation*}
\widehat{S}_{\text{ph}}\Phi _{M}^{\hbar }=\Phi _{\alpha (S)M}^{\hbar }\text{
\ , \ }\widehat{S}_{\text{ph}}\Phi _{z_{0},M}^{\hbar }=\widehat{T}_{\text{ph}%
}(Sz_{0})\widehat{S}_{\text{ph}}\Phi _{\alpha (S)M}^{\hbar }\text{.}
\end{equation*}

\subsection{Nearby-orbit method in phase space}

Let us now state and prove the main result of this paper:

\begin{proposition}
(i) Let $U(t,z_{0})$ be the semiclassical propagator for Schr\"{o}dinger's
equation and set $\psi =U(t,z_{0})\psi _{0}$. The wavepacket transform $%
U_{\phi }$ takes $\psi $ to the function $\Psi $ defined by%
\begin{equation}
\Psi =e^{\frac{i}{\hbar }\gamma (z_{0},t)}\widehat{T}_{\text{ph}%
}(f_{t}(z_{0}))(\widehat{S}_{t}(z_{0}))_{\text{ph}}\widehat{T}(z_{0})_{\text{%
ph}}^{-1}\Psi _{0}  \label{propa}
\end{equation}%
with $\Psi _{0}=U_{\phi }\psi _{0}$. (ii) In particular, if $\psi _{0}=\phi
_{z_{0}}^{\hbar }$ 
\begin{equation}
U_{\phi }(U(t,z_{0})\phi _{z_{0}}^{\hbar })=e^{\frac{i}{\hbar }\gamma
(z_{0},t)}\widehat{T}_{\text{ph}}(f_{t}(z_{0}))(\widehat{S}_{t}(z_{0}))_{%
\text{ph}}\Phi _{0}^{\hbar }.  \label{propb}
\end{equation}
\end{proposition}

\begin{proof}
Set $\psi =U(t,z_{0})\psi _{0}$; by definition of $U(t,z_{0})$ we have%
\begin{equation*}
\psi =e^{\frac{i}{\hbar }\gamma (z_{0},t)}\widehat{T}(f_{t}(z_{0}))\widehat{S%
}_{t}(z_{0})\widehat{T}(z_{0})^{-1}\psi _{0}
\end{equation*}%
hence, using successively (\ref{14}) and (\ref{16}), 
\begin{eqnarray*}
U_{\phi }\psi &=&e^{\frac{i}{\hbar }\gamma (z_{0},t)}U_{\phi }(\widehat{T}%
(f_{t}(z_{0}))\widehat{S}_{t}(z_{0})\widehat{T}(z_{0})^{-1}\psi _{0} \\
&=&e^{\frac{i}{\hbar }\gamma (z_{0},t)}\widehat{T}_{\text{ph}}(f_{t}(z_{0}))(%
\widehat{S}_{t}(z_{0}))_{\text{ph}}\widehat{T}(z_{0})_{\text{ph}%
}^{-1}U_{\phi }\psi _{0}.
\end{eqnarray*}%
Formula (\ref{propb}) follows since we have%
\begin{equation*}
\widehat{T}(z_{0})_{\text{ph}}^{-1}\Phi _{z_{0}}^{\hbar }=\widehat{T}%
(z_{0})_{\text{ph}}^{-1}U_{\phi }(\phi _{z_{0}}^{\hbar })=U_{\phi }(\widehat{%
T}(z_{0})^{-1}\phi _{z_{0}}^{\hbar })=U_{\phi }(\phi ^{\hbar })
\end{equation*}%
and $\Phi _{0}^{\hbar }=U_{\phi }(\phi ^{\hbar })$.
\end{proof}

The result above therefore suggests that the phase-space version of the
semiclassical nearby-orbit propagator $U(t,z_{0})$ should be given by the
formula

\begin{equation}
U_{\text{ph}}(t,z_{0})=e^{\frac{i}{\hbar }\gamma (z_{0},t)}\widehat{T}_{%
\text{ph}}(f_{t}(z_{0}))(\widehat{S}_{t}(z_{0}))_{\text{ph}}\widehat{T}%
(z_{0})_{\text{ph}}^{-1}  \label{phpr}
\end{equation}%
and we have:

\begin{proposition}
Let $\Psi $ be the solution of the phase-space Schr\"{o}dinger equation 
\begin{equation*}
i\hslash \frac{\partial \Psi }{\partial t}=\widehat{H}_{\text{ph}}\Psi \text{
\ , \ }\Psi (t=0)=\Phi _{z_{0}}^{\hbar }
\end{equation*}%
with $\Phi _{z_{0}}^{\hbar }=U_{\phi }(\phi _{z_{0}}^{\hbar })$. Suppose
that $H$ satisfies the conditions (\ref{condh}) in Proposition \ref{probou}.
Then, for $|t|<T$ there exists a constant $C_{T}\geq 0$ such that 
\begin{equation}
||U_{\text{ph}}(t,z_{0})\Phi _{z_{0}}^{\hbar }-\Psi (\cdot ,t)||_{L^{2}(%
\mathbb{R}_{z}^{2n})}\leq C_{T}\sqrt{\hbar }.  \label{ineq}
\end{equation}
\end{proposition}

\begin{proof}
The inequality (\ref{ineq}) is an easy consequence of the estimate (\ref%
{estim}): the solution $\Psi $ is given by $\Psi (\cdot ,t)=U_{\phi }(\psi
(.,t))$ where $\psi $ is the solution of the usual Schr\"{o}dinger equation%
\begin{equation*}
i\hslash \frac{\partial \psi }{\partial t}=\widehat{H}\psi \ ,\ \ \psi
(t=0)=\phi _{z_{0}}^{\hbar };
\end{equation*}%
since $U_{\phi }$ is a linear isometry we have%
\begin{equation*}
||U_{\text{ph}}(t,z_{0})\Phi _{z_{0}}^{\hbar }-\Psi (\cdot ,t)||_{L^{2}(%
\mathbb{R}_{z}^{2n})}=||U(t,z_{0})\phi _{z_{0}}^{\hbar }-\psi (\cdot
,t)||_{L^{2}(\mathbb{R}_{x}^{n})}\leq C_{T}\sqrt{\hbar }.
\end{equation*}
\end{proof}

\section{Conclusion and Discussion}

There are several problems and questions we have not discussed in this
paper, and to which we will come back in forthcoming publications. Needless
to say, there is one outstanding omission: we haven't analyzed the domain of
validity of the nearby-orbit method much in detail. There are many results
in the literature for the standard nearby-orbit method. For instance,
Hagedorn \cite{Hag1,Hag2} obtains precise estimates using the Lie--Trotter
formula; similar results were rediscovered and sharpened by
Combescure-Robert \cite{CR1} using the Duhamel principle. Also see \cite%
{Naza} (Ch.2, \S 2.1). It shouldn't be too difficult to obtain corresponding
estimate for the phase-space Schr\"{o}dinger equation, using the properties
of the wavepacket transform. It is on the other hand well-known that there
are problems with long times when the associated classical systems exhibits
a chaotic behavior; as Littlejohn points out in \cite{Littlejohn}, the
nearby orbit methods probably fails for long times near classically unstable
points.

\begin{acknowledgement}
This work has been supported by the FAPESP grant 2005/51766--7 during the
author's stay at the University of S\~{a}o Paulo. I take the opportunity to
thank Professor Paolo Piccione for his kind and generous invitation.
\end{acknowledgement}


\begin{thebibliography}{99}
\bibitem{Bast} M J Bastiaans 1979 Wigner distribution function and its
application to first order optics. \textit{J. Opt. Soc. Am} \textbf{69}
1710--1716.

\bibitem{CR} M Combescure and D Robert 2005 Quadratic Quantum Hamiltonians
revisited. math-ph/0509027.

\bibitem{CR1} M Combescure and D Robert 1997 Semiclassical spreading of
quantum wave packets and applications near unstable fixed points of the
classical flow. \textit{Asymptotic Analysis} \textbf{14} 377--404

\bibitem{Folland} G B Folland 1989 Harmonic Analysis in Phase space. \textit{%
Annals of Mathematics studies, Princeton University Press, Princeton}, N.J.

\bibitem{AIF} M de Gosson 1990 Maslov Indices on $\limfunc{Mp}(n)$.\textit{\
Ann. Inst. Fourier,} Grenoble, \textbf{40}(3) 537--555.

\bibitem{MdGMW} M de Gosson 2005 The Weyl Representation of Metaplectic
Operators. \textit{Lett. Math. Phys}, \textbf{72} 129--142.

\bibitem{phystv} M de Gosson 2005 Extended Weyl Calculus and Application to
the Phase-Space Schr\"{o}dinger Equation. \textit{J. Phys. A: Math. and
General} \textbf{38}.

\bibitem{physa} M de Gosson 2005 Symplectically Covariant Schr\"{o}dinger
Equation in Phase Space. \textit{J. Phys.A:Math. Gen}. \textbf{38}
9263--9287.

\bibitem{Birk} M de Gosson 2006 Symplectic Geometry and Quantum Mechanics. 
\textit{Birkh\"{a}user, Basel,} Series Operator Algebras and Applications.

\bibitem{Hag1} G Hagedorn 1981 Semiclassical quantum mechanics III. \textit{%
Ann. Phys.} \textbf{135}, 58--70

\bibitem{Hag2} G Hagedorn 1985 Semiclassical quantum mechanics IV. \textit{%
Ann. Inst.} H. Poincar\'{e} \textbf{42}, 363--374.

\bibitem{Haas} M Haas 1994 Siegel's Modular Forms and Dirichlet Series. 
\textit{Lecture Notes in Mathematics} \textbf{216}, Springer--Verlag, 1971.

\bibitem{Heller} E Heller 1975 Time-dependent approach to semiclassical
dynamics. \textit{The Journal of Chemical Physics} \textbf{62}(4),1544--1555.

\bibitem{Basil} B Hiley 2006 Moyal's Characteristic Function, the Density
Matrix and von Neumann's Idempotent. Preprint, University of London.

\bibitem{KS} J R Klauder and E C G Sudarshan 1968 Fundamentals of Quantum
Optics. Benjamin, New York.

\bibitem{Leray} J Leray 1981 Lagrangian Analysis and Quantum Mechanics,\ a
mathematical structure related to asymptotic expansions and the Maslov
index. \textit{The MIT Press}, Cambridge, Mass.

\bibitem{Littlejohn} R G Littlejohn 1986 The semiclassical evolution of wave
packets. \textit{Physics Reports} \textbf{138(}4--5), 193--291.

\bibitem{MF} V P Maslov and M V Fedoriuk 1981 \textit{Semi-Classical
Approximations in Quantum Mechanics}. Reidel, Boston.

\bibitem{Naza} V Nazaikiinskii, B-W Schulze, and B Sternin 2002 Quantization
Methods in Differential Equations. Differential and Integral Equations and
Their Applications, \textit{Taylor \& Francis}.

\bibitem{Pere} A Perelomov 1986 \textit{Generalized Coherent States and
their Applications}. Springer-Verlag, Berlin, Heidelberg.

\bibitem{TV1} G Torres-Vega and J H Frederick 1990 Quantum mechanics in
phase space: New approaches to the correspondence principle. \textit{J.
Chem. Phys}. \textbf{93}(12), 8862--8874.

\bibitem{TV2} G Torres-Vega and J H Frederick 1993 A quantum mechanical
representation in phase space. \textit{J.} \textit{Chem. Phys.} \textbf{98}%
(4), 3103--3120.

\bibitem{Wallach} N Wallach 1977 \textit{Lie Groups: History, Frontiers and
Applications}, \textbf{5}. Symplectic Geometry and Fourier Analysis, Math
Sci Press, Brookline, MA.
\end{thebibliography}
\end{document}